\documentclass{emulateapj}
\usepackage{apjfonts}
\usepackage{epsf}

\begin{document}

\title{LOW-METALLICITY STAR FORMATION IN HIGH-REDSHIFT GALAXIES
       AT $z \sim$  8}

\author{
Y. Taniguchi,\altaffilmark{1} 
Y. Shioya,\altaffilmark{1} and 
J. R. Trump\altaffilmark{2}
}

\altaffiltext{1}{Research Center for Space and Cosmic Evolution, Ehime University, 
        Bunkyo-cho, Matsuyama 790-8577, Japan}
\altaffiltext{2}{Steward Observatory, University of Arizona, Tucson, AZ 85721}

\shortauthors{Taniguchi et al.}
\shorttitle{Star Formation at Redshift of 8}

\begin{abstract}
Based on the recent very deep near-infrared imaging of the Hubble Ultra Deep Field 
with WFC3 on the {\it Hubble Space Telescope}, five groups published most probable samples 
of galaxies at $z \sim 8$, selected by the so-called dropout method or 
photometric redshift; e.g., $Y_{105}$-dropouts ($Y_{105}-J_{125} > 0.8$). 
These studies are highly useful for investigating both the early star formation history 
of galaxies and the sources of cosmic re-ionization. 
In order to better understand these issues, 
we carefully examine if there are low-$z$ interlopers 
in the samples of $z \sim 8$ galaxy candidates. 
We focus on the strong emission-line galaxies at $z \sim 2$ in this paper. 
Such galaxies may be selected as $Y_{105}$-dropouts 
since the [O{\sc iii}] $\lambda$ 5007 emission line is redshifted into the $J_{125}$-band. 
We have found that the contamination from such low-$z$ interlopers is
negligibly small. Therefore, all objects found by the five groups are 
free from this type of contamination.
However, it remains difficult to extract real $z \sim 8$ galaxies because 
all the sources are very faint and the different groups have found different candidates. 
With this in mind,
we construct a robust sample of eight galaxies at
 $z \sim 8$ from the objects
found by the five groups:
each of these eight objects has been selected by at least two groups.
Using this sample, we discuss their UV continuum slope. 
We also discuss the escape fraction of ionizing photons adopting various metallicities. 
Our analysis suggests that 
massive stars forming in low-metallicity gas 
($Z \sim 5 \times 10^{-4} Z_{\odot}$) 
can be responsible for the completion of cosmic re-ionization if
the escape fraction of ionizing continuum from galaxies is as 
large as 0.5, and this is consistent with the observed blue UV continua.
\end{abstract}



\section{INTRODUCTION}

In order to understand both the early phase of galaxy evolution and 
the cosmic history of intergalactic space 
(i.e., the cosmic re-ionization of intergalactic medium), 
it is crucially important to probe any objects at very high redshift. 
In this decade, ground-based 8-m class telescopes have enabled us to find a large number 
of galaxies beyond $z \sim 6$ (e.g., Hu et al. 2002; Rhoads et al. 2004; 
Taniguchi et al. 2005; Stern et al. 2005; Kashikawa et al. 2006; Iye et al. 2006; 
Ota et al. 2008; Ouchi et al. 2009; see for a review Taniguchi 2008). 
Also, the optical deep imaging survey promoted with the {\em Hubble Space Telescope} 
({\em HST}), the Hubble Ultra Deep Field (HUDF; Beckwith et al. 2006), resulted 
in very probable detection of galaxies beyond $z = 7$ together with deep near-infrared 
(NIR) imaging surveys with {\em HST} and large ground-based telescope facilities 
(Bouwens et al. 2006, 2008).
In addition, recent deep near infrared imaging surveys 
with 8-m class ground-based telescopes
provided probable candidates of galaxies beyond $z=7$ (e.g., Hibon et al. 2010).

More recently, the successful installation of the Wide Field Camera 3 (WFC3) on 
{\em HST} was promptly used to observe HUDF in the NIR window. 
Using these imaging data, very promising candidates of high-$z$ galaxies 
at $z \sim 7$ -- 10 were found by several researchers 
(Bouwens et  al. 2009, 2010b; Oesch et al. 2010;
Bunker et al. 2010; McLure et al. 2010; Yan et al. 2010; Finkelstein et al. 2010).
These objects were used to investigate both the early star formation history
in galaxies and the cosmic re-ionization history.
In fact, the following new results were obtained in these studies.
 (1) The number density of luminous galaxies appears to become smaller 
with increasing redshift while 
that of less luminous ones appears to be constant 
(Bouwens et al. 2010b). 
(2) The galaxies beyond $z \sim 7$ tend to have bluer UV continuum
with $\beta < -3$ 
($f_\lambda \propto \lambda^\beta$), significantly bluer than
the typical $\beta \sim -2$ of nearby starburst galaxies and Lyman break galaxies (LBGs) at $z < 6$  (Bouwens et al. 2010a, 2010b; Finkelstein et al. 2010).
This trend is more pronounced for fainter galaxies. 
And, (3) the escape fraction of ionizing UV continuum is also discussed
in terms of the cosmic re-ionization at $z > 7$.
For example, if more numerous, low-luminosity star-forming galaxies
could be present, they would be enough to reionize the intergalactic
medium together with both a larger escape fraction of ionizing
photons from them and a small clumping factor of the universe
 (e.g., Yan et al. 2010; Bunker et al. 2010).
However, there appears no consensus among the studies.

Any conclusions and suggestions derived through these studies depend strongly
on the robustness of the detection of such very high-$z$ galaxies. 
In order to examine how the results are robust, 
it is important to reject any contamination from low-$z$ interlopers.
Here, we focus on galaxies at $z \sim 8$ (i.e., $Y_{105}$ dropouts)
because the following five groups extracted their own samples of
galaxies at $z \sim 8$ based on the WFC3 observations of the HUDF;
Bouwens et al. (2010b), Bunker et al. (2010), McLure et al. (2010), 
Yan et al. (2010), and Finkelstein et al. (2010). 

Possible interlopers are low-temperature stars, low-$z$ red galaxies,
and low-$z$ very strong emission-line galaxies with faint continua.
All groups but Bunker et al. (2010) removed compact objects (i.e., possible low-temperature stars in our Galaxy) from
their samples. They also checked the possibility of low-$z$ red galaxies
by using {\em Spitzer} IRAC photometry and confirmed that such contamination
cannot occur in their analysis.
However, they did not examine the possibility of contamination from
strong emission-line galaxies at low $z$, although Oesch et al. (2010) pointed out 
such possibility only for $z \sim 7$ ($z_{850}$-dropout) galaxies.
Therefore, in this paper, we first examine whether or not this type of 
contamination occurs in their analysis.

Then, we construct a robust sample of galaxies at $z \sim 8$
free from any contamination. In this procedure, we also take account of
possible false detection due to unexpected photometric effects
discussed by Bouwens et al. (2009).
Based on this new sample together with the results of Bouwens et al. 
(2010b, 2010c), we discuss the early star formation history
of galaxies and cosmic re-ionization at $z \sim 8$.

Throughout this paper, magnitudes are given in the AB system. 
We adopt a flat universe with $\Omega_{\rm matter}=0.3$, $\Omega_\Lambda=0.7$, 
and $H_0=70 \; {\rm km \; s^{-1} \; Mpc^{-1}}$. 

\section{THE SELECTION CRITERIA FOR IDENTIFYING GALAXIES AT $z \sim$ 8}

Before studying the possibility of contamination by strong emission-line galaxies
at $z \sim 1.5$ -- 2, it is necessary to examine how the galaxies
at $z \sim 8$ have been identified in HUDF by the following five groups: 
(1) Bouwens et al. (2010b), (2) Bunker et al. (2010), (3) McLure et al. (2010), 
(4) Yan et al. (2010), and (5) Finkelstein et al. (2010). 
In this section, 
we summarize their selection criteria for identifying galaxies at $z \sim 8$.

First, we list up all the candidates identified by the five groups in Table 1.
The numbers of detected galaxies at $z \sim 8$ are 5, 7, 8, 15, and 9 
in Bouwens et al. (2010b),  Bunker et al. (2010), McLure et al. (2010), 
Yan et al. (2010), and Finkelstein et al. (2010), respectively. 
Their studies lead to 20  galaxies at $z \sim 8$ in total.
Note that the last object (ID No. 21) is selected as a $z_{850}$ dropout 
in Oesch et al. (2010), Bunker et al. (2010), and Yan et al. (2010)
while as a $z \sim 8$ galaxy candidate in McLure et al. (2010) and 
Finkelstein et al. (2010). 
We treat this object as a galaxy at $z \sim 7$ and thus we do not use it
in this paper. 
Yan et al. (2010) identified their z8-SD15 as YD5 in Bunker et al. (2010).
However, checking their positions and thumbnails, we identify z8-SD24 as YD5 and
there is no counterpart of z8-SD15 in the samples of the other three groups. 

Only four galaxies (Nos. 1, 2, 3, and 4 in Table 1)
are commonly identified by all five groups.  
Since the five groups made their own data reduction,
their final images are different from each other. 
In fact, their zero points are slightly different from each other.
In addition, their selection criteria for galaxies at $z \sim 8$ 
are also different; see a summary of their selection criteria
given in Table \ref{tab:selection}.
Three groups (Bouwens et al. 2010b; Bunker et al. 2010; Yan et al. 2010) 
selected $z \sim 8$ galaxies using the dropout method ($Y_{105}$-dropout) 
while the others (McLure et al. 2010; Finkelstein et al. 2010) selected 
$z \sim 8$ galaxies using the photometric redshift technique. 
Although the selection criteria are slightly different from group
to group, we adopt 
$Y_{105}-J_{125} > 0.8$ as our primary criterion for galaxies at $z \sim 8$.

\section{STRONG EMISSION-LINE GALAXIES AS INTERLOPERS}

Possible interlopers are very strong emission-line galaxies
with very weak continuum emission like blue compact dwarf galaxies;
e.g., I Zw 18 (Izotov et al. 1999). 
Strong emission lines may modify observed broadband colors and 
thus make false continuum break features\footnote{Here, 
we do not consider active galactic nuclei (AGNs).
Although AGNs are also strong emission-line sources, they have brighter 
rest-frame blue continuum and thus they may be easily detected in the very
deep ACS imaging of HUDF. Therefore, we focus on star-forming galaxies like
blue compact dwarf galaxies in the local universe.}.

Actually, in the local universe, metal-poor galaxies 
with large emission-line equivalent widths have been surveyed using their unusual 
broadband colors caused by strong emission lines 
(Brown et al. 2008). 
Recently, using the data of the Sloan Digital Sky Survey Data Release 7 
(Abazajian et al. 2009), Cardamone et al. (2009) investigated a class of 
compact star-forming galaxies called ``Green Peas.'' 
Because of their very strong [O{\sc iii}]$\lambda$5007 emission line with
a rest-frame equivalent width of $\rm EW_0$([O{\sc iii}])$>500$ \AA, 
``Green Peas'' in the redshift range of $0.112 < z < 0.360$ are very bright in $r$ band. 
Their $r-i$ colors are bluer by  $>0.5$ mag than those of normal galaxies. 
Such ultra strong emission-line galaxies have also been found up to $z \sim 1$ 
(Kakazu et al. 2007; Hu et al. 2009; see also Ohyama et al. 1999). 

\begin{figure}
\epsscale{0.5}
\plotone{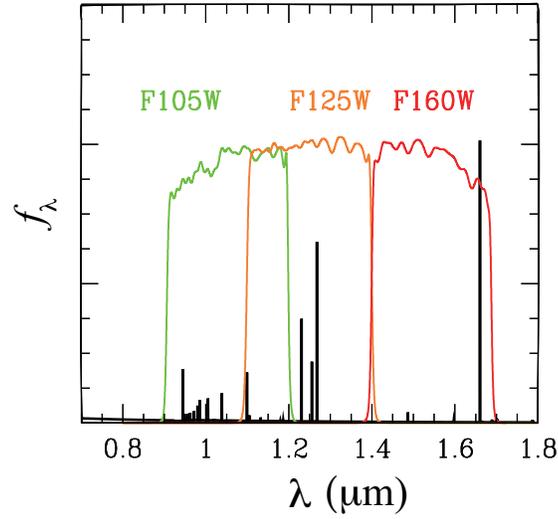}
\caption{
Example of the SED of a strong emission-line galaxy at $z=1.5$.
This spectrum corresponds to one model spectrum generated in Section 4;
the SED with age of 100 Myr. Note that this is also similar to that 
of I Zw 18 (Izotov et al. 1999). 
The filter responses of $F105W$, $F125W$, and $F160W$ are also shown. 
\label{fig:emline_zwl}
}
\end{figure}

Since the strongest emission line is 
generally [O {\sc iii}]$\lambda$5007
for such dwarf starburst galaxies 
(e.g., Izotov et al. 1999; Brown et al. 2008; Cardamone et al. 2009), 
it is possible that strong emission-line galaxies at $z \sim 1.5$ -- 2
could be detected as $Y_{105}$ dropouts if their optical continuum emission
is very weak. To demonstrate this, we show a typical spectrum of a strong 
emission-line galaxy at $z = 1.5$ (Figure \ref{fig:emline_zwl}).
Therefore, in this paper, 
we first examine if this type of contamination could accidentally occur
in the identification of galaxies at $z \sim 8$ in HUDF.

\section{MODELS OF LOW-$z$ STAR-FORMING GALAXIES}

We make our model spectral energy distributions (SEDs) for star-forming 
galaxies.
First, we generate model galaxy SEDs with nebular continuum emission 
using the population synthesis model, STARBURST99 (Leitherer et al. 1999), 
adopting the mode of constant star formation (SFR) rate together with
both Salpeter initial mass function
(IMF; the lower and upper masses are 1 $M_{\odot}$ and 100 $M_{\odot}$, respectively)
and the metallicity of $Z=0.2Z_\odot=0.004$. 
These models correspond to nearby blue compact dwarf galaxies
with very high SFRs and sub-solar metallicities
(e.g., $Z \sim 0.1 Z_\odot$), such as I Zw 18 (Izotov et al. 1999). 

The nebular continuum is important in this analysis 
since it affects the global shape of the SEDs, 
especially for young starburst galaxies (Schaerer 2003; Bouwens et al. 2010a). 
We calculate the luminosity of the nebular continuum (free-free,
free-bound, and two photon emission for H and He)
assuming that the escape fraction of the ionizing continuum is $f_{\rm esc} = 0$. 
This value is consistent with observations of star-forming galaxies
at $z \sim 1$ -- 3 (e.g., Malkan et al. 2003; Siana et al. 2007; 
Steidel et al. 2001); where we expect strong emission-line
galaxies as possible interlopers.

However, since nebular emission lines are not included in the above
SEDs generated by STARBURST99, 
we add some emission lines to the SED generated above. 
Our procedures are as follows.
First, we calculate the luminosity of H$\beta$ emission line, 
$L({\rm H}\beta)$, using the following relation (Leitherer \& Heckman 1995):
\begin{equation}
L({\rm H}\beta) \; ({\rm erg \; s^{-1}}) = 4.76 \times 10^{-13}N_{\rm Lyc}\; ({\rm s^{-1}}), 
\end{equation}
where $N_{\rm Lyc}$ is the ionizing photon production rate
estimated in each SED model. 
Second, we estimate other emission lines
(e.g., [O {\sc ii}] $\lambda$ 3727, [O {\sc iii}] $\lambda$ 5007,
and so on),
adopting the observed emission-line ratios of I Zw 18 (Izotov et al. 1999),  
which is one of the most metal-poor galaxies known
in the local universe. 
Then, we add all emission lines with $F(\lambda)/F({\rm H}\beta) > 0.004$ 
from [O {\sc ii}] $\lambda$ 3727 to [Ar {\sc iii}] $\lambda$ 7135. 
Note that the emission-line equivalent widths tend to be larger
with decreasing metallicity because here we use the fixed emission-line ratios
of I Zw 18.

\begin{figure}
\epsscale{0.7}
\plotone{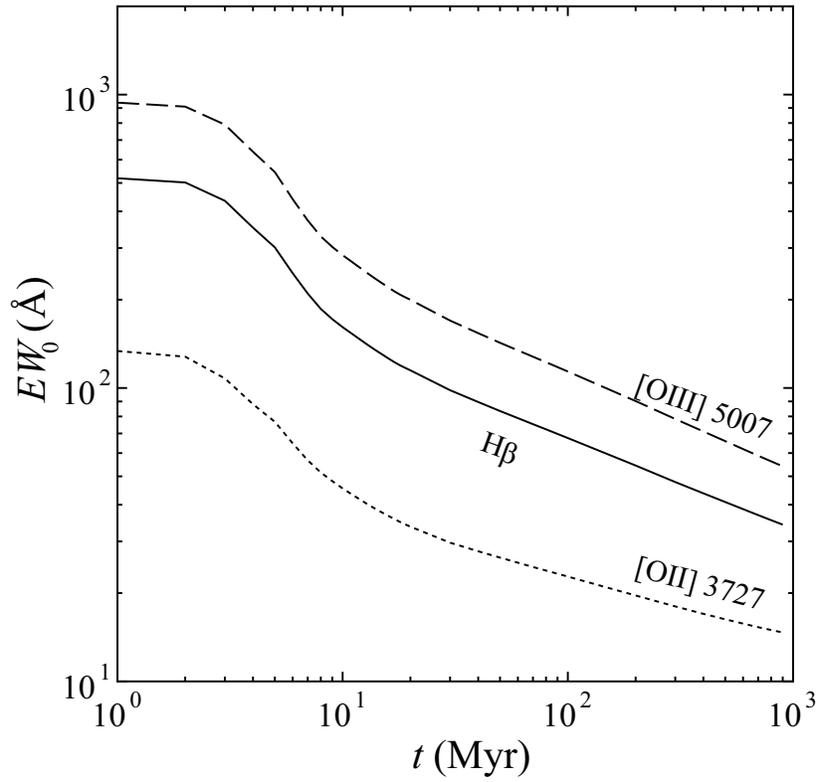}
\caption{
Evolution of $\rm EW_0$(H$\beta$), $\rm EW_0$([O {\sc ii}]$\lambda$3727),
and $\rm EW_0$([O{ \sc iii}]$\lambda$5007), 
for the constant star formation model with Salpeter IMF and $Z=0.2 Z_{\odot}=0.004$. 
\label{fig:ew_csfe2_2.eps}
}
\end{figure}

Given the star formation history, the equivalent widths of emission lines
vary with the age of a galaxy because the emission-line luminosities
are proportional to the production rate of ionizing photons. 
Figure \ref{fig:ew_csfe2_2.eps} shows the evolution of equivalent width of 
H$\beta$, [O {\sc iii}]$\lambda$5007, and [O {\sc ii}]$\lambda$3727 
in our model SEDs. 
In the next section, we examine the variation of observed colors as a function 
of redshift for the model galaxies with ages of 1, 10, and 100 Myr. 
The values of $\rm EW_0$(H$\beta$) are 520, 160, and 70 \AA~ 
for galaxies with ages of 1, 10, and 100 Myr,
respectively. 

\section{RESULTS}

\begin{figure}
\epsscale{0.7}
\plotone{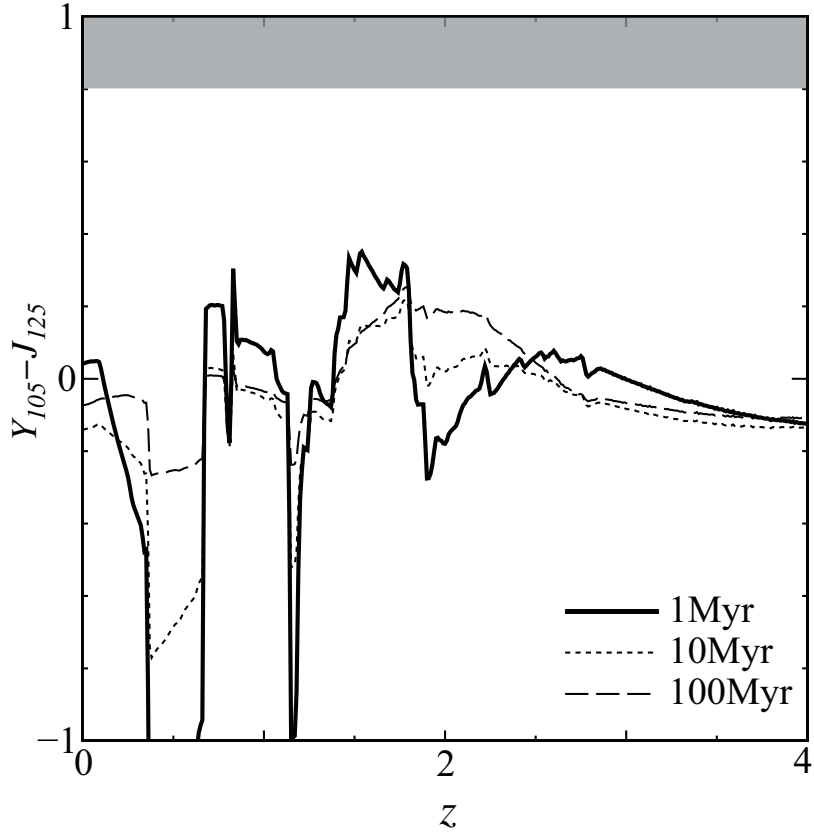}
\caption{
Expected color of $Y_{105}-J_{125}$ is shown as a function of
redshift for model galaxies with ages of 1 Myr (solid curve), 10 Myr (dotted curve), 
and 100 Myr (dashed curve). The shaded area shows the primary selection criterion
of $Y_{105} - J_{125} > 0.8$.
\label{fig:colorvsz}
}
\end{figure}

First, we examine if some low-$z$ strong emission-line galaxies accidentally 
satisfy the selection criterion for galaxies at $z \sim 8$ galaxies; i.e.,
$Y_{105}-J_{125} > 0.8$.
In Figure \ref{fig:colorvsz}, we show the variation of $Y_{105}-J_{125}$ colors of the model galaxies with age of 1, 10, and 100 Myr described in Section 4 
as a function of redshift ($0 \le z \le 4$).
It is shown that the reddest $Y_{105}-J_{125}$ color is achieved
at $1.47 \le z \le 1.81$ for the model with age of 1 Myr. 
In this case, the $J_{125}$ magnitude is dominated by
the four strong emission lines of
[O {\sc iii}]$\lambda$5007, [O {\sc iii}]$\lambda$4959, H$\beta$, 
and H$\gamma$ (see Figure \ref{fig:emline_zwl}).
However, the color is as red as $Y_{105}-J_{125} \simeq 0.4$
because [O {\sc ii}]$\lambda$3727 contributes to the flux in $Y_{105}$.
Another red peak with  $Y_{105}-J_{125} \simeq 0.35$
 is seen at $z \simeq 0.83$. This feature is due to
redshifted H$\alpha$ emission in $J_{125}$.
A broad red peak around $z \sim 2.5$ is due to [O {\sc ii}]$\lambda$3727
in $ J_{125}$, but the color is only $Y_{105}-J_{125} \sim 0.1$.
Here we note that the [O {\sc ii}] emission is generally weaker than [O {\sc iii}]
in metal-poor, star-forming galaxies (e.g., Nagao et al. 2006).
For the models with ages of 10 Myr and 100 Myr, the reddest color
appears to be $Y_{105}-J_{125} \sim$ 0.2 -- 0.25 at most around $z \sim 1.8$.

In summary, strong emission-line galaxies at $1.47 \le z \le 1.81$
give rise to $Y_{105}-J_{125} \sim 0.3$ with the reddest color of
0.35 at $z \sim 1.54$. These colors are much bluer than the
primary criterion of $Y_{105}-J_{125} > 0.8$ (e.g., Bouwens et al. 2010b).
However, at this stage, we cannot conclude that there is no contamination by
strong emission-line galaxies around $z \sim 1.5$ because the following
two points must be also taken into account; (1) reddening, and 
(2) photometric errors in the WFC3 observations of HUDF.

As for the reddening, it is known that nearby blue compact dwarf galaxies
are reddened in the range of  $0 < A_V < 1$ (Kong et al. 2003).
However, we do not know how star-forming galaxies at $z \sim$ 2 
galaxies are typically reddened. 
Since they may be more massive than nearby blue compact dwarf galaxies, 
they may be more strongly reddened.
For example, the extinction in star-forming galaxies at $z \sim 2$ 
selected as BX galaxies ranges from $A_V \simeq 0$ to 1.5 (Shapley et al. 2005). 
Taking account of this, for the safety, we investigate the three cases; 
$A_V=0$, 1, and 2 mag together with 
the reddening curve for starburst galaxies obtained by Calzetti et al. (2000). 
To examine the reddening effect, first we use 
the $Y_{105}-J_{125}$ versus $J_{125}-H_{160}$ diagram
because this diagram is used to identify galaxies at $z \sim 8$
(e.g., Bouwens et al. 2010b). 
In addition, we also use another diagram of 
$Y_{105}-J_{125}$ versus $i_{775}-J_{125}$ diagram.
The reason is as follows. If some $z \sim 8$ galaxy candidates
are strong emission-line galaxies at $z \sim 2$, 
the optical Advanced Camera for Surveys (ACS) imaging data of the HUDF are also useful.
Since the ACS survey depth (5$\sigma$, $0.35^{\prime\prime} ~ \phi$)
is 29.4 AB for $B_{435}$, 29.8 AB for $V_{606}$,
29.7 AB for $i_{775}$, and 29.0AB for $z_{850}$ (Bouwens et al. 2010b),
$i_{775}-J_{125}$ gives the strongest constraint on 
the contamination possibility; see Appendix A.
Following the selection criteria of $z \sim 8$ galaxies in the five groups,
we calculate a lower limit of $i_{775}-J_{125}$ color 
adopting $2 \sigma$ upper limit of $i_{775}$. 

\begin{figure}
\epsscale{1.0}
\plotone{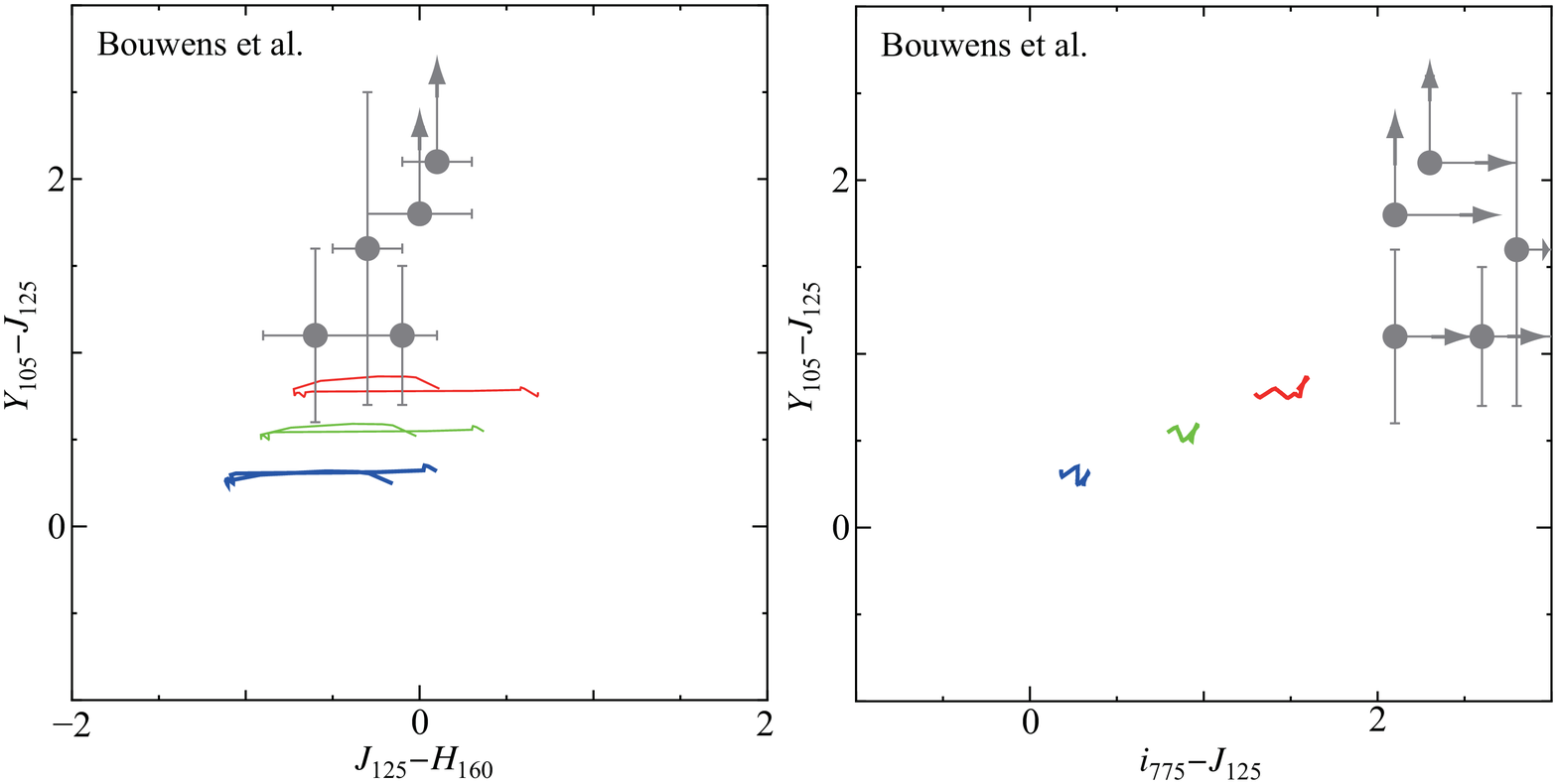}

\caption{
Diagrams of $Y_{105}-J_{125}$ vs. $J_{125}-H_{160}$ (left), and 
$Y_{105}-J_{125}$ vs. $i_{775}-J_{125}$ (right). 
Blue, green, and red curves show 
the loci of strong emission-line galaxies between $z=1.47$ and $z=1.81$ 
with $A_V=0$, 1, and 2 mag. 
The observational data of 
Bouwens et al. (2010b) (a), 
Bunker et al. (2010) (b), 
McLure et al. (2010) (c), 
Yan et al. (2010) (d), and 
Finkelstein et al. (2010) (e) 
are shown as filled gray circles with error bars.
\label{fig:YJJH}
}
\end{figure}

In Figure \ref{fig:YJJH}, we show comparisons between our SED models with the effect of
reddening and the WFC3 photometry by the five groups.
Since we also show the photometric errors in each diagram,
we can examine the possibility of contamination from
strong emission-line galaxies at $z$ = 1.47 to 1.81 
taking account of both the effects of reddening and photometric errors
simultaneously.

It is extremely difficult to estimate the properties of the population
of potential contaminant $1.5 \lesssim z \lesssim 2$ emission line
galaxies, but we can qualitatively estimate the robustness of each $z
\sim 8$ candidate using the reddening required for an emission-line
galaxy at $1.5 \lesssim z \lesssim 2$ to mimic that object's colors.
At first, we summarize our results as follows.
 (1) The $z \sim 8$
galaxy candidates identified both by Bouwens et al. (2010b) and 
by Bunker et al. (2010) are free from the contamination of
strong emission-line galaxies unless strong emission-line
galaxies with $A_V > 2$ exist at  $z \sim$ 1.5 -- 2. 
(2) However, if strong emission-line galaxies
with $A_V =$ 1 -- 2 are present at  $z \sim$ 1.5 -- 2, 
it is possible that some of
the $z \sim 8$ galaxy candidates identified by McLure et al. (2010),
Yan et al. (2010), and Finkelstein et al. (2010) are strong emission-line
galaxies at $z \sim$ 1.5 -- 2.

Here we should note that the above different results among the five groups
are due to the differences in photometry; i.e.,  some candidates have 
a non-zero probability to be an interloper in the sample of one group,
but not another. However, more importantly, we should also mention that
heavily obscured dwarf galaxies with very strong emission lines
are probably rare from observational material obtained to date. 
Therefore, we conclude that the possibilities of low-$z$ interlopers are 
negligibly small for all the $z \sim 8$ galaxies 
identified by each group.

\section{DISCUSSION}

\subsection{A sample of robust galaxies at $z \sim 8$ }

In this section, we discuss the properties of galaxies robustly determined to be at $z \sim 8$. 
In the last section, we have shown that the contamination of 
low-$z$ strong emission-line galaxies is negligibly small
for all candidates studied by the five groups. 
However, it is important to note that there might be other unexpected
contamination sources  (see, e.g., the summary of Bouwens et al. 2009),
and so careful procedures are absolutely necessary to identify robust
candidates at high redshift even if we use deep WFC3 imaging data.
In particular, they find that a large fraction of nine candidates 
(10 -- 18 in Table \ref{tab:candidats}) identified only by Yan et al. (2010) 
are quite close to foreground galaxies and they may be
affected by foreground bright sources. 
We therefore do not use the nine objects. 
We also do not use both 19 and 20 that are detected only by Finkelstein et al. (2010) since these objects are also close to foreground bright objects.
Taking all these into account, we have constructed a sample of
eight robust galaxies at $z \sim 8$ all that are
identified by at least two groups. These galaxies are from
1 to 8 in Table 1.
We summarize their photometric properties in Table 3. 
We note here that our robust sample is merely robust in the sense that these candidates 
have been selected by at least two groups and do not lie near foreground galaxies: spectroscopic follow-up is necessary to confirm that they are 
truly robust $z > 8$ galaxies. 

Here it is noted that all the five groups tried to do their best
in extraction of the total magnitude for their sample galaxies although
their methods and photometric zero points are slightly 
different\footnote{The photometry methods of the five groups are
summarized below;  (1) Bouwens et al. (2010b): MAG\_AUTO with Kron factor = 1.2 
and then corrected to the case of Kron factor of 2.5 (this corresponds to 
           the so-called aperture correction),
  (2) Bunker et al. (2010): aperture photometry with 0.6 arcsec $\phi$ and
           then applied an aperture correction,
  (3) McLure et al. (2010): aperture photometry with 0.6 arcsec $\phi$ and
           then applied an aperture correction,
  (4) Yan et al. (2010): MAG\_AUTO with Kron factor = 1.2 and then applied 
           an aperture correction, and
  (5) Finkelstein et al. (2010): MAG\_AUTO with Kron factor = 1.2 
           and then applied an aperture correction.
}.
In fact, the magnitudes and colors for the same objects tabulated
in Table 3 are slightly different among the groups.
Therefore, in our later analysis (Section 6.3),
we use mean values given in the last column of Table 3.
For the value of $x \pm \sigma$, we calculate the average of $x$ ($\langle x \rangle$), 
the systematic error of $x$ ($\sigma_{\rm sys}$), 
the mean of the random error $\sigma_{\rm rdm}$, 
and the total errors $\sigma_{\rm tot}$ are as follows: 
$\langle x \rangle = (\sum_{i=1}^N x_i)/N$, 
$\sigma_{\rm sys} = \sqrt{(1/N)\sum_{i=1}^N(x_i - \langle x \rangle)^2}$, 
$\sigma_{\rm rdm} = \sqrt{(\sum_{i=1}^N \sigma_i^2)}/N$, and 
$\sigma_{\rm tot} = \sqrt{ \sigma_{\rm sys}^2 + \sigma_{\rm rdm}^2}$, 
where $N$ is the number of data. 
The lower limit values are simply ignored for calculating the mean photometry,
although in all cases they are consistent with the mean value from the true detection.

\subsection{Comment on the cosmic re-ionization}

In this section, we discuss how the star-forming galaxies contribute to the cosmic re-ionization at $z \sim 8$. 
Since the cosmic reioniztion is one of important issues in high-redshift universe, several investigations have 
been made for these years.
Based on the survey of $z$-band dropouts in the Subaru Deep Field and the Subaru XMM-Newton Deep Survey, 
Ouchi et al. (2009) evaluated the ionizing photon production rate of star-forming galaxies at $z \sim 7$ 
adopting Salpeter stellar initial mass function (IMF) and sub-solar metallicity ($Z=0.2Z_\odot$). Comparing it 
with the critical ionizing photon production rate obtained by Madau et al. (1999), they found the ionizing photon 
production rate of star-forming galaxies is enough to complete re-ionization of the universe if the escape fraction 
is larger than 0.2 and suggested that the properties of very high redshift ($z \sim 7$) galaxies are different from 
those at low redshifts, e.g., a larger escape fraction, a lower metallicity, and/or a flatter initial mass function 
if the very high redshift universe is ionized by only galaxies. 
Bunker et al. (2010) and McLure et al. (2010) also discussed the relation between the high-$z$ galaxies and 
the cosmic re-ionization at $z \sim 7$ based on the survey of $z$-band dropouts in the Hubble Ultra Deep Field.
Bunker et al. (2010) pointed out that the ionizing photon production rate of star-forming galaxies at $z \sim 7$ 
is lower than the critical value adopting the relation for the solar metallicity. 
McLure et al. (2010) calculated the evolution of the filling factor of the ionized hydrogen assuming the sub-solar 
metallicity ($Z=0.2Z_\odot$) and the escape fraction of 0.2 and found that the inter-galactic medium would not 
achieve a unity HII filling factor until $z = 4.2$. 
Their findings also suggested that the star-forming galaxies at very high-$z$ are the large escape fraction of 
ionizing photons, the lower metallicity, and/or the top-heavy IMF.

Now, we focus on the cosmic re-ionization by galaxies at $z \sim 8$. 
Here we use the formalism studied by Madau et al. (1999).
However, we do not use the critical SFR to complete
the cosmic re-ionization (e.g., Equation (27) in Madau et al. 1999)
since the Salpeter IMF and the solar metallicity are assumed there.
Instead, we estimate the production rate of ionizing photons
based on the UV luminosity functions obtained by Bouwens et al. (2010b, 2010c). 
Using Equation (26) in Madau et al. (1999) together with the update of 
cosmological parameters given in Section 1, we obtain the following  
critical ionizing photon production rate density,

\begin{equation}
\dot{\mathcal{N}}_{\rm ion} = \frac{1.2 \times 10^{51}{\rm photons \; s^{-1} \; Mpc^{-3}}}{f_{\rm esc}} 
\left( \frac{1+z}{9} \right)^3 
\left( \frac{\Omega_{\rm b} h_{70}^2}{0.0463} \right)^2
\left( \frac{C}{5} \right)
\end{equation}
where $\Omega_{\rm b}$ is the cosmic baryon density and 
$h_{70}$ is the Hubble parameter in units of $h=0.7$. 
$C$ is the clumping factor of neutral hydrogen, 
$C = \langle \rho_{\rm HI}^2 \rangle \langle \rho_{\rm HI} \rangle^{-2}$, 
where $\rho_{\rm HI}$ and $\langle \rho_{\rm HI} \rangle$ are the local and 
cosmic mean density of neutral hydrogen (Gnedin \& Ostriker 1997). 
We adopt $\Omega_{\rm b} h_{70}^2 =0.0463$ 
from {\em Wilkinson Microwave Anisotropy Probe} 5 Year data (Komatsu et al. 2009). 
The clumping factor $C$ is conventionally assumed to be 30 
based on cosmological radiative transfer simulations by 
Gnedin \& Ostriker (1997). 
However, smaller values for $C$ have been recently suggested by
cosmological hydrodynamic simulation in which photo-ionization heating 
together with radiative cooling 
is taken into account (Pawlik et al. 2009).
Since the pressure support increases and then smooths out small-scale
density fluctuations, the photo-ionization heating reduces the 
clumping factor down to $C \sim 3$ -- 5. The clumping factor 
depends on the starting redshift of reheating, $z_{\rm r}$.
If $z_{\rm r}$ = 19.5, $C \simeq 3$, while 
if $z_{\rm r}$ = 9, $C \simeq 5$.
In this paper, we adopt $C =5$.
Therefore, the production rate of ionizing photons
is determined by only one parameter, $f_{\rm esc}$, 
that is the average escape fraction of hydrogen-ionizing photons
from galaxies to intergalactic space. 

\begin{figure}
\epsscale{0.8}
\plotone{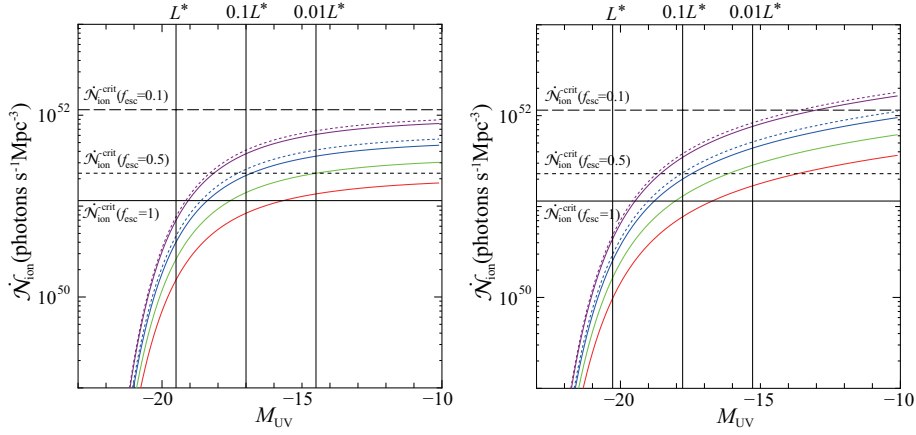}
\caption{
Cumulative ionizing photon production rate density 
($\dot{\mathcal{N}}_{\rm ion}$ ) is shown as a function of $M_{\rm UV}$ 
for the UV luminosity function of Bouwens et al. (2010b) (left panel) and 
Bouwens et al. (2010c) (right panel).
Solid purple, blue, green, and red curves correspond to the models with $Z=0$, 
$5 \times 10^{-4}Z_\odot$, $0.02Z_\odot$, and $1 Z_\odot$, respectively,
with the Salpeter IMF of 
($M_{\rm lower}$, $M_{\rm upper}$)=($1 M_\odot$, $100 M_\odot$).
Dotted purple and blue curves correspond to the models
with $Z=0$ and $5 \times 10^{-4}Z_\odot$, with the Salpeter IMF of 
($M_{\rm lower}$, $M_{\rm upper}$)=($1 M_\odot$, $500 M_\odot$).
Three vertical lines show the critical ionizing photon production rate density 
that is necessary to keep the universe 
completely ionized at $z = 8$: the line for $f_{\rm esc}=1$, the
dotted line for $f_{\rm esc}=0.5$, and the dashed line for $f_{\rm esc}=0.1$. 
\label{fig:sfrd}
}
\end{figure}

In Figure. \ref{fig:sfrd}, we compare the critical ionizing photon production rate density 
with the cumulative ionizing photon production rate densities corresponding to the UV 
luminosity function derived by Bouwens et al. (2010b, 2010c). 
The UV luminosity function of Bouwens et al. (2010b) was obtained by using their 
five  $Y_{105}$-dropout 
galaxies in HUDF discussed in this paper. 
The best-fit Schechter function parameters are 
$M^*=-19.5 \pm 0.3$, $\phi^*=1.1 \times 10^{-3}{\rm Mpc^{-3}}$ (fixed), and $\alpha = -1.74$ (fixed). 
Recently, Bouwens et al. (2010c) obtained 
a new UV luminosity function for galaxies at $z \sim 8$ using 47 $Y_{105}$ or $Y_{098}$-dropout galaxies
found in the additional two HUDF09 fields and deep, wide-area Early Release Science (ERS). 
\footnote{The selection criteria in Bouwens et al. (2010c) are slightly different from those 
in Bouwens et al. (2010b): they adopted $Y_{105}-J_{125} > 0.45 \wedge 
J_{125}-H_{160} < 0.5$ for HUDF09, HUDF09-1, and HUDF09-02. Since the ERS observations use 
a different $Y$-band filter ($Y_{098}$), the selection criteria for ERS are 
$Y_{098}-J_{125} > 1.25 \wedge J_{125}-H_{160} < 0.5$.} 
The best-fit Schechter function parameter set is 
$M^*=-20.28 \pm 0.19$, $\phi^*=0.38^{+0.57}_{-0.22} \times 10^{-3}{\rm Mpc^{-3}}$, 
and $\alpha = -2.00 \pm 0.33$. 
Because the faint-end slope of the UV luminosity function of Bouwens et al. (2010c) 
is steeper than that of Bouwens et al. (2010b), 
the contribution of faint galaxies to the ionizing photon production rate density 
is larger for the UV luminosity function of Bouwens et al. (2010c) than 
for Bouwens et al. (2010b). 

The ratio of ionizing photon production rate to 
UV (rest-frame 1500 \AA) luminosity depends on the metallicity.
Since it is probable that the metallicity of stars formed in 
the galaxies at $z \sim 8$ is lower than the solar value,
we investigate the following cases; $Z = 0$ (i.e., the Population III
condition), $Z = 5 \times 10^{-4} Z_{\odot}$,
$Z = 0.02 Z_{\odot}$, and $Z = 1 Z_{\odot}$.
In order to convert the UV luminosity density to 
ionizing photon production rate density, we adopt the relation between them 
at age of 10 Myr\footnote{It has been often discussed that the ages of 
the high-z galaxies are estimated as  a few $\times$ 100 Myr or shorter (e.g.,
Finkelstein et al. 2010; Labbe et al. 2010). 
Since the mean stellar mass that ionizing photons is more massive than 
that emit UV continuum, the ratio of $Q_{\rm H}/L_{\rm UV}$ becomes smaller 
with increasing age.
If we evaluate the time averaged $Q_{\rm H}/L_{\rm UV}$
between $t=0$ and 100 Myr,
the ratio of $Q_{\rm H}/L_{\rm UV}$ becomes $1/2$ and $2/3$ of the value at $t$=10Myr,
respectively. 
This trend is found not only for the case of solar metallicity but also for
 those of lower metallicity discussed in this paper.}
for the constant star formation model with Salpeter IMF 
of ($M_{\rm lower}$, $M_{\rm upper}$) =($1 M_{\odot}$, $100 M_{\odot}$) 
calculated by Schaerer (2003). In addition, we also adopt another 
Salpeter IMF of  ($M_{\rm lower}$, $M_{\rm upper}$) =($1 M_{\odot}$, $500 M_{\odot}$) for $Z = 0$ and $Z = 5 \times 10^{-4} Z_{\odot}$.
The basic properties used in this analysis are given in Appendix B;
i.e., the evolution of ionizing photon production rate, $Q_{\rm H}$,
rest-frame UV luminosity at 1500 \AA, $L_\lambda$(1500),
and their ratio,  $Q_{\rm H} / L_\lambda$(1500) as a function of age.
We note here that we do not include any correction for dust obscuration. 

The escape fraction required to reionize the universe is summarized in Table \ref{tab:fesc} 
for the various metallicity and upper mass cutoff mentioned above.
The tabulated values of $f_{\rm esc}$ are shown as a function of $L^*$ 
(i.e., stars that contribute to the ionizing photons). 
It is here noted that the stellar mass of an $L^*$ galaxy is 
estimated as $M_{\rm star} \sim 3.5 \times 10^8 M_{\odot}$ for the UV luminosity function of Bouwens et al. (2010b) 
\footnote{We use the relation of $M_{\rm star} / L_{\rm UV} \simeq 10^{-32}
M_{\odot} / ({\rm erg \; s^{-1}}$\AA$^{-1}$) for the constant
star formation model by Schaerer (2003). This relation is valid 
for age with older than 100 Myr. }. 
Note also that Labbe et al. (2010) obtained $M^* \sim 10^9 M_{\odot}$
for three galaxies at $z \sim 8$ in HUDF from their 
very deep $Spitzer$/IRAC observation.

It is shown from Table \ref{tab:fesc} that galaxies at $z \sim 8$ with the
solar metallicity can ionize the intergalactic medium
if less-massive objects with $M_{\rm UV} \sim -15$ also
form massive stars.
Note, however, that such objects have only $M_{\rm star} \sim 
10^7 M_{\odot}$. 
Moreover, in order to keep the universe ionized, 
the source-averaged escape fraction must be as high as $f_{\rm esc} \sim 0.9$ 
for the UV luminosity function of Bouwens et al. (2010b) 
and $\sim 0.7$ for that of Bouwens et al. (2010c). 
It seems likely that some galaxies have dusty clouds because
supernova events could enrich heavy elements in the interstellar
gas and then cause the formation of dust grains. 
Since dusty clouds make the escape fraction smaller
(e.g., Yajima et al. 2009), the source-averaged escape fraction
could be close to unity.
Therefore, in the case of solar metallicity,
the cosmic re-ionization cannot be solely achieved 
by star-forming galaxies at $z \sim 8$.

The chemical evolution of theoretical models of galaxies
suggests that the metallicity of galaxies at $z \sim 8$
is less than $0.01 Z_{\odot}$ 
(e.g., Razoumov \& Sommer-Larsen 2010).
It is then more reasonable to investigate 
the escape fraction in the low-metallicity conditions.
In the case of $Z \sim 0.02 Z_{\odot}$,
galaxies brighter than $0.1 L^*$ 
could reionize the universe for both UV luminosity function. 
Under the assumption that galaxies are the only ionizing sources, 
their source-averaged escape fraction would have to be as large as ~0.9.
On the other hand,
in the case of $Z = 5 \times 10^{-4} Z_{\odot}$, 
we can obtain a more modest required escape fraction,
$f_{\rm esc} \sim 0.5$ for ionization by galaxies with
$L > 0.1 L^*$ ($M_{\rm star} > 10^8 M_{\odot}$).

Larger escape fractions (up to
$f_{\rm esc} \sim 0.8$) at higher redshift
(e.g., $z \sim 4$ - 10) are suggested from 
theoretical points of view (Wise \& Cen 2009;
Razoumov \& Sommer-Larsen 2010).
Yajima et al. (2009) investigated the escape fraction for
Ly$\alpha$ emitters (LAEs) and LBGs at $z \sim$ 
3 -- 7, taking account of both the effect of dust extinction
and collisional ionization by superwind shocks.
They found $f_{\rm esc} \sim$ 0.07 -- 0.47 for LAEs
and $f_{\rm esc} \sim$ 0.06 -- 0.17 for LBGs.
Although the escape fraction depends on the systemic mass of 
galaxies, the lower metallicity condition (i.e.,
$Z = 5 \times 10^{-4} Z_{\odot}$) seems to be more reasonable.

Finally, we comment on the evolution of
the escape fraction from very high redshift (i.e., $z \sim 8$)
to the present day. The available observations 
indicate $f_{\rm esc} < 0.01$ for $z < 1$ while
$f_{\rm esc} \sim 0.01$ -- 0.1 for $z \sim 1$ -- 3
(Inoue et al. 2006 and references therein,
Steidel et al. 2001; Shapley et al. 2006; Siana et al. 2007;
Iwata et al. 2009 and references therein).
However, if galaxies were the only sources responsible
for re-ionization, $f_{\rm esc} \sim 0.5$ would be required.
This dramatic evolution can be understood if 
mass assembly processes could be working at $z \sim 8$.
In this evolutionary phase, typical masses of galaxies 
are less massive (i.e., $< 10^9 M_{\odot}$)
than those in the present day, the negative feedback from 
supernovae is more effective and thus superwind-driven bubbles 
help the escape of ionizing photons from galaxies at $z \sim 8$ 
(Mori \& Umemura 2006; Wise \& Cen 2009).

\subsection{Blue color of UV continuum}

We discuss the origin of very blue color of UV continuum 
found in some galaxies at $z \sim 8$;
i.e., $\beta < -3$ ($f_\lambda \propto \lambda^\beta$).
This property has been first found in galaxies at $z \sim 7$
(Bouwens et al. 2010a; Finkelstein et al. 2010).
Note that local starbursts and star-forming galaxies 
at $z \sim 5$ - 6 show $\beta \sim -2$.
Then, Bouwens et al. (2010b) also noted that their galaxies
at $z \sim 8$ tend to have similar properties (see also Finkelstein et al. 2010).
Since the Ly$\alpha$ emission falls in $J_{125}$ band
for the galaxies at $z \sim 8$, more careful check is necessary 
to investigate their UV continuum.
In order to investigate this issue,
we discuss effects of the escape fraction of ionizing continuum 
and Ly$\alpha$ emission shifted to the $J_{125}$ window.

\begin{figure}
\epsscale{0.8}
\plotone{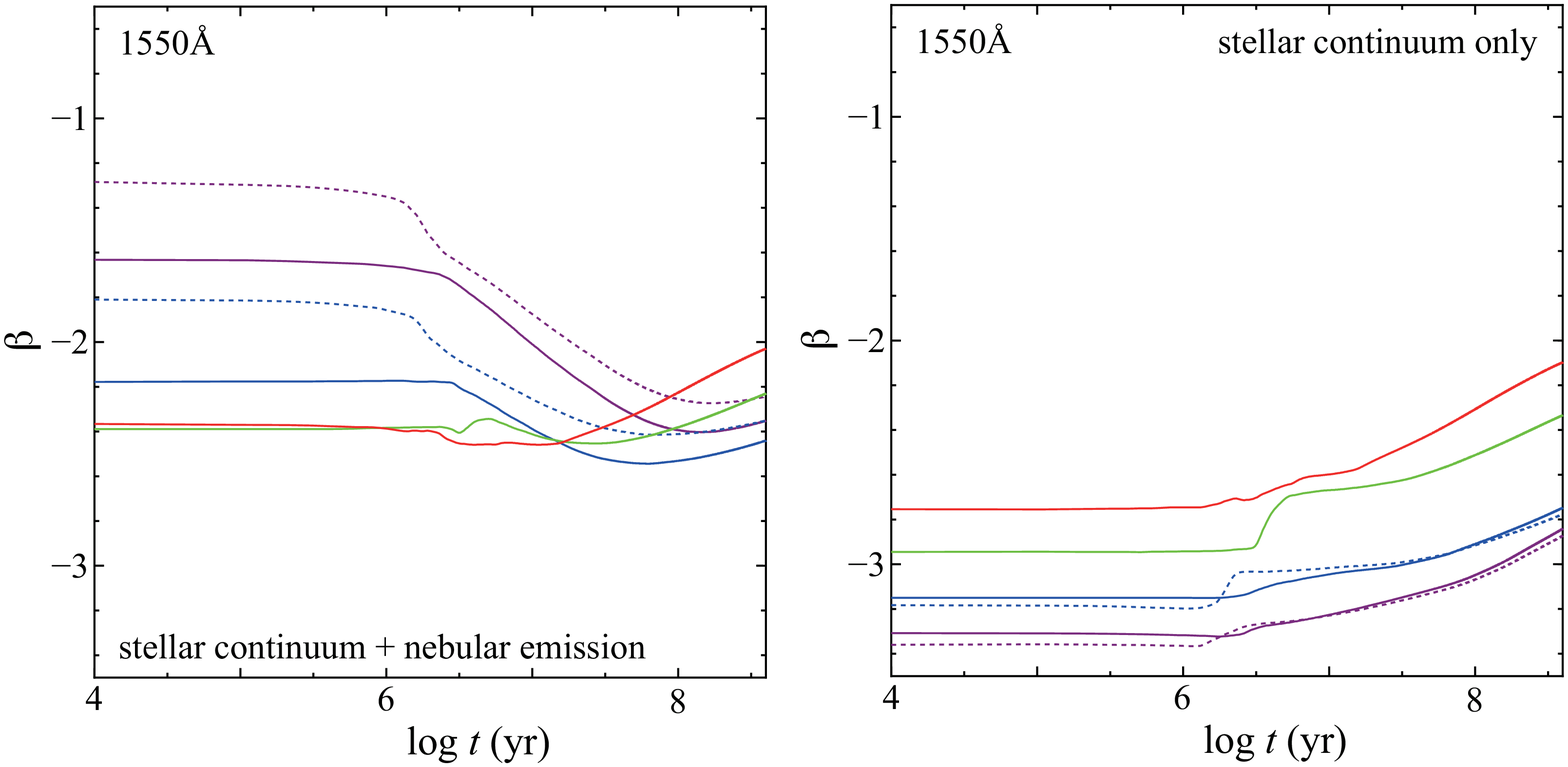}
\caption{
Evolution of UV continuum slope, $\beta$, 1550 \AA ~ in the rest frame
is shown as a function of age of the system for different metallicities.
These model results are taken from
the constant star formation models by Schaerer (2003): 
 (left) stellar continuum + nebular emission, and (right) stellar continuum only. 
The meaning of each curve is the same as that in Figure \ref{fig:sfrd}.
\label{fig:betaage}
}
\end{figure}

Based on the population synthesis model of Schaerer (2003), 
the UV continuum slope $\beta$ is shallower than $-2.5$ 
for any metallicity if the nebular continuum emission is 
taken into account. 
However, in the case of no nebular continuum emission, 
$\beta$ becomes steeper than $-3$ if the metallicity is lower than
$Z < 5 \times 10^{-4} Z_{\odot}$. 
These properties are shown in Figure \ref{fig:betaage};
see also Bouwens et al. (2010a).

If the escape fraction is small (i.e., $f_{\rm esc} \sim 0$),
the ionizing continuum is absorbed by H {\sc i} gas within the galaxy,
causing galaxy-scale H {\sc ii} regions. As a result,
the nebular continuum emission is so strong that the color of 
UV continuum becomes redder; e.g., $\beta > -2.5$.
In the case of $Z \sim 5 \times 10^{-4} Z_{\odot}$,
the escape fraction would have to be $f_{\rm esc} \sim 0.5$ 
if galaxies with $L > 0.1 L^*$ were to keep the universe 
ionized  (see Table \ref{tab:fesc}).
If this is the case,
the nebular continuum emission is highly suppressed and thus 
blue colors of UV continuum should be observed.

\begin{figure}
\epsscale{0.8}
\plotone{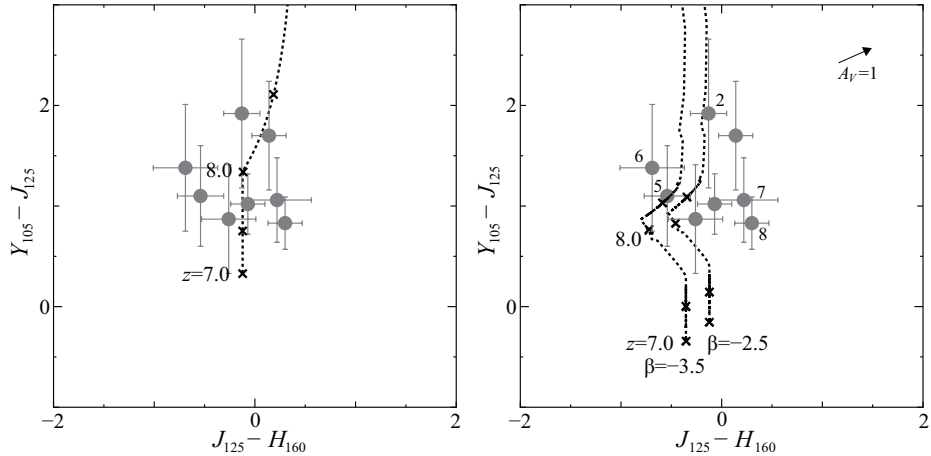}
\caption{Comparison between the observed colors and model results in
$Y_{105}-J_{125}$ vs. $J_{125}-H_{160}$ diagram.
Left: observed colors are shown together with 
the locus for model galaxies with $\beta = -2.5$ at $z > 7$ (dotted curve).
Right: observed colors are shown together with 
the loci for model galaxies with $\beta = -2.5$ and 
$\beta = -3.5$ at $z > 7$ (dotted curve). 
The flux contribution of Ly$\alpha$ emission with  $\rm EW_0({\rm Ly}\alpha)=250$\AA~
is also taken into account for both models.
The reddening vector for $A_V=1$ mag is at shown the upper-right corner.
\label{fig:yjjh}
}
\end{figure}

In Figure \ref{fig:yjjh}, we compare the observed colors and the model results.
For simplicity, the model loci (dotted curve) for $\beta = - 2.5$ 
and $-3.5$ are shown as a function of redshift from $z=7$.
Here we use the averaged photometry given in Table 3. 
All galaxies appear to be consistent with $\beta = -2.5$ at the $2\sigma$ level,
especially if we consider reddening as low as $A_V \sim 1$.
However, the observed scatter in this Figure appears to be large
(see also Bouwens et al. 2010b; Finkelstein et al. 2010).
Since the age of galaxies at $z \sim 8$ may be from several Myr to 
a few hundred Myr (note that the age of the universe at $z \sim 8$ is
600 Myr), the observed color distributions are understood if 
the nebular continuum emission is negligibly small (see the right panel
of Figure \ref{fig:betaage}).

Two galaxies (5 and 6) appear to have much bluer 
UV continuum; i.e., $\beta \sim -4$. These colors cannot be 
explained even with very young, extremely metal poor 
stellar populations (see again the right panel of Figure \ref{fig:betaage}).
One possibility is that strong Ly$\alpha$ emission 
contributes to the $J_{125}$ band flux.
In order to explain the observed blue $J_{125} - H_{160}$ colors,
the equivalent width of Ly$\alpha$ emission must be 
as large as 250 \AA.
In the right panel of Figure \ref{fig:yjjh}, we show the loci of model galaxies with
Ly$\alpha$ emission line [$\rm EW_0({\rm Ly}\alpha)=250$ \AA]
for $\beta = -2.5$  and  $\beta = -3.5$. It is shown that 
the observed data of 5 and 6 are explained with the effect
of Ly$\alpha$ emission.
Some high-$z$ Ly$\alpha$ emitters have $\rm EW_0({\rm Ly}\alpha)$ 
larger than 100 \AA ~ (e.g., Nagao et al. 2004, 2005, 2007; 
Murayama et al. 2007; Shioya et al. 2009). 
We note that a large amount of intergalactic neutral hydrogen absorbs a large fraction 
of the Ly$\alpha$ photons, which would reduce the impact of Ly$\alpha$ on the colors. 
If the metallicity of the galaxies at $z \sim 8$ is much 
smaller than that at $z \sim 5$ -- 6, the Ly$\alpha$ equivalent 
width is expected to be much larger than  100 \AA; e.g., 
$\rm EW_0({\rm Ly}\alpha) >$ 500 \AA ~ (Schaerer 2003).

Another interesting property found in Figure \ref{fig:yjjh} is that 
the other two galaxies, 7 and 8, show redder colors,
corresponding to $\beta \sim -1$. If we explain this property
by dust extinction, the extinction would have to be as large as $A_V \sim 2$,
corresponding to the rest-frame UV extinction, $A_{\rm UV} \sim 5$.
If this is the case, these galaxies are brighter by 5 mag than
the estimated absolute magnitudes; i.e., they would be as bright as 
$M_{\rm UV} \sim -24$ mag.
Yajima et al. (2009) pointed out that the escape fraction of 
ionizing photons 
is reduced by dust extinction in a factor of from a few to 10.
For dusty galaxies, the escape fraction of the ionizing photons becomes small. 
If these two galaxies are highly extincted and have resultantly low escape fractions, 
then to achieve the average escape fraction of 
$f_{\rm esc} \sim 0.5$, 
the escape fraction of other galaxies at $z \sim 8$ 
must be nearly unity. 
If the cosmic re-ionization is completed by star-forming galaxies,
it is again suggested that lower metallicity conditions
are more favorable at $z \sim 8$.

\section{CONCLUDING REMARKS}

In this paper, we have found that strong emission-line galaxies
at $z \sim 1.5$ -- 2 do not work as interlopers for the identification of
the 20 galaxies at $z \sim 8$ in the HUDF by the five groups
(Bouwens et al. 2010b; Bunker et al. 2010; McLure et al. 2010;
Yan et al. 2010; Finkelstein et al. 2010). 
Although they are free from low-$z$ interlopers, 
there may be other unexpected problems in the procedures in the source detection.
Taking these points into account, 
we have constructed a robust sample of eight galaxies at $z \sim 8$ that are all identified 
by at least two groups.
Using this sample, we have investigated the observed bluer color of UV continuum. 
Based on the two UV luminosity functions obtained by Bouwens et al. (2010b, 2010c), 
we have also investigated the contribution of the galaxies at $z \sim 8$ 
to the cosmic re-ionization together with the escape fraction of ionizing 
continuum from the galaxies. Our results and conclusions are summarized below.

\begin{enumerate}
\renewcommand{\labelenumi}{\arabic{enumi}.}

\item Based on the UV luminosity functions of Bouwens et al. (2010b, 2010c), 
we estimate the ionizing photon production rates for various stellar populations with
metallicity from $Z = 0$ (i.e., the Population III condition)
to $Z = Z_{\odot}$. Comparing these rates to the critical values 
for the completion of cosmic re-ionization, we have found 
for the solar metallicity stars that 
the escape fraction must be close to unity even if we take 
account of ionizing photons from 
galaxies with $\sim 0.01 L^*$.
On the other hand, in the case of low metallicity condition,
$Z \sim 5 \times 10^{-4} Z_{\odot}$, the modest escape fraction,
$f_{\rm esc} \sim 0.5$, from galaxies with $L > 0.1 L^*$
is required to achieve the cosmic re-ionization.
We therefore suggest that metal-poor conditions, e.g.,
$Z \sim 5 \times 10^{-4} Z_{\odot}$, 
are more favorable for star formation in galaxies at $z \sim 8$.

\item Such a large escape fraction can be responsible for the observed very blue
UV continuum color, $\beta < -3$, for some galaxies at $z \sim 8$.
It is also expected that a large number of Ly$\alpha$ photons can escape from
galaxies and then contribute to the flux in $J_{125}$ band. 
These two factors are responsible for the observed blue colors 
of UV continuum.

\end{enumerate}

In summary, the stellar populations of the galaxies at $z \sim 8$ 
are dominated by significantly low-metallicity stars
with $Z \sim 5 \times 10^{-4} Z_{\odot}$. Subsequent explosions of supernovae
lead to multiple superwinds  in the universe at $z \sim 8$, 
and then re-ionize the universe.
All the observational properties of the galaxies at $z \sim 8$ 
appear to be consistent with this scenario.

In this paper, we assume that low-metallicity stars resided in 
the galaxies at $z \sim 8$ would keep the universe ionized, 
there may be some other ionization sources at such high redshift;
e.g., pure Population III stars (Choudhury \& Ferrara 2006), 
kinematic heating (collisional ionization) by supernaovae
(Tegmark et al. 1993; see also Miniati et al. 2004), 
mini quasars (Madau et al. 2004), and so on.
However, we are not able to determine
which sources are more dominant for the cosmic re-ionizations at $z \sim 8$. 
Future observational and theoretical investigations will be necessary
to resolve this important issue.

\vspace{1pc}
We thank Tohru Nagao for useful discussion.
We also thank the anonymous referee for comments and
suggestions that have improved the clarity of this paper.
This work was financially supported in part by the JSPS 
(Nos. 17253001 and 19340046).
J.R.T. acknowledges support from NSF/DDEP grant 0943995.


\appendix

\section{Optical dropout constraints to reject the contamination by strong 
Emission-line galaxies at $z \sim 1.5$ to 2}
\label{sec:whyij}

\begin{figure}
\epsscale{0.8}
\plotone{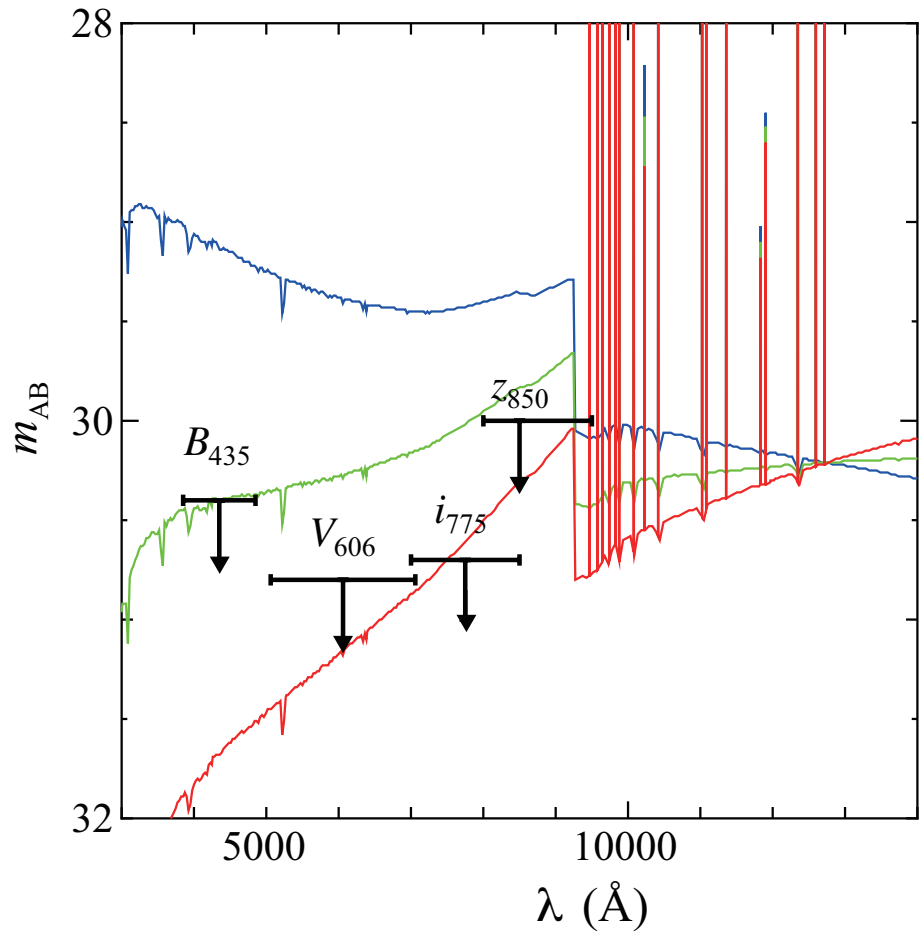}
\caption{
Limiting magnitudes for the ACS filter bands in the HUDF (Bouwens et al.
2010b) are shown together with the three SEDs of a star-forming galaxy at
$z =1.5$; $A_V =0$ (blue), $A_V =1$ (green), and $A_V =2$ (red).
\label{fig:SEDvsACSlim}
}
\end{figure}

We demonstrate here that the constraint from the lower limit 
of $i_{775}-J_{125}$ color 
is strongest among the upper limit of $B_{435}-J_{125}$, $V_{606}-J_{125}$, $i_{775}-J_{125}$, 
and $z_{850}-J_{125}$ (i.e., non detection of all optical ACS bands). 
Figure \ref{fig:SEDvsACSlim} shows the relation between the upper limits of ACS bands and

In Figure \ref{fig:SEDvsACSlim}, we show the three SEDs of a strong emission-line galaxy 
of $J=29.2$ at $z = 1.5$ for $A_V=0$, 1, and 2. 
The detection limits of the ACS imaging of HUDF are also shown
in this figure (Bouwens et al. 2010b).
As shown here, if such strong emission-line galaxies are dust free, 
they must be detected in all ACS bands. 
However, in the case of $A_V=2$,
the predicted magnitudes of the galaxy are fainter than the upper limits in 
$B_{435}$, $V_{606}$, and $z_{850}$ bands while
the predicted magnitude of $i_{775}$ band is brighter than the upper limit.
Therefore, the lower limit of $i_{775}-J_{125}$ color 
is strongest among the upper limit of $B_{435}-J_{125}$, $V_{606}-J_{125}$, $i_{775}-J_{125}$, and $z_{850}-J_{125}$.
The magnitude of the model SED in Figure \ref{fig:SEDvsACSlim} ($J=29.2$) may seem fainter than 
the observed galaxy. We note that the limiting magnitude shown in this figure is 
measured with $0.35^{\prime \prime}$ diameter aperture. 
On the other hand, the magnitudes of galaxies at $z \sim 8$ are the total ones
(see the footnote in Section 6.1). 

\section{Metallicity dependence of the ionizing continuum}
\label{sec:NionZ}

Many authors (e.g., Bunker et al. 2010) have compared the SFR density 
based on the UV luminosity density with the critical SFR density 
formulated  by Madau et al. (1999). 
In their formulation, the Salpeter IMF with 
($M_{\rm lower}$, $M_{\rm upper}$)=($1 M_\odot$, $100 M_\odot$)
and the solar metallicity are used. 
However, it is known that
the ionizing photon production rate and the UV luminosity depend strongly on 
the metallicity (e.g., Schaerer 2003).
In order to discuss star formation properties in high-$z$ galaxies,
it seems better to investigate star formation in metal poor environs.

\begin{figure}
\epsscale{0.4}
\plotone{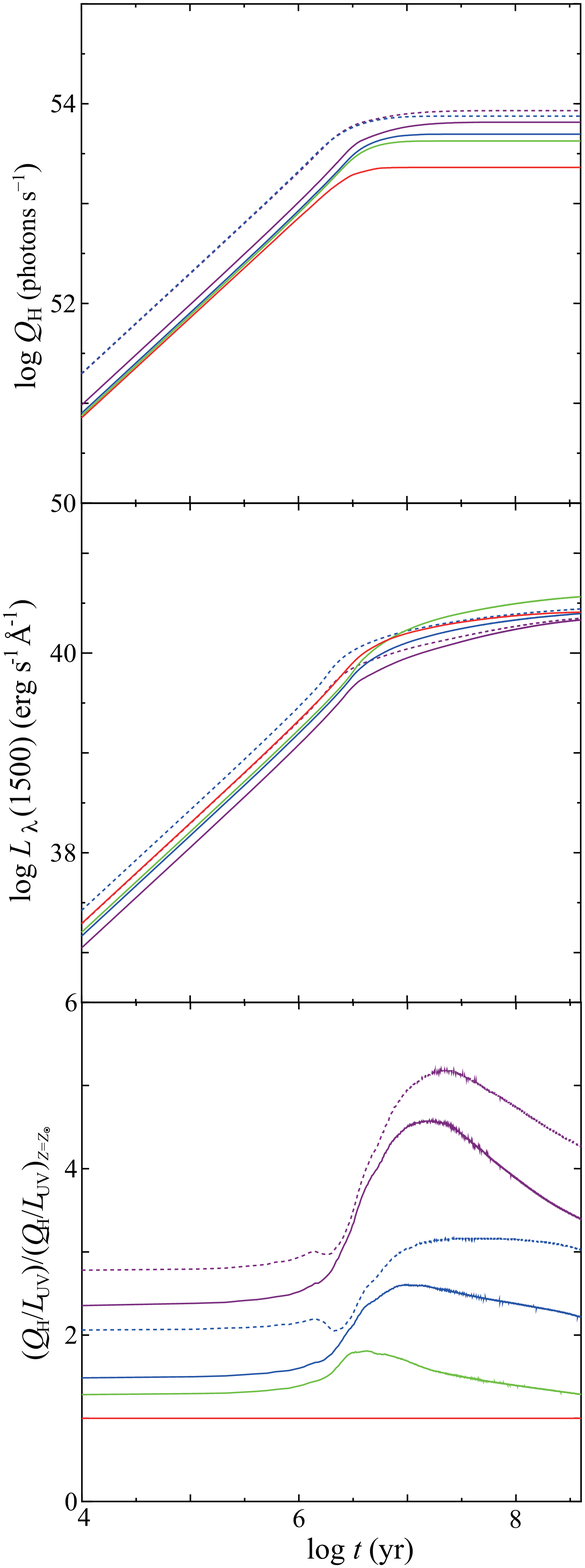}
\caption{
Upper panel: evolution of ionizing photon production rate, $Q_{\rm H}$. 
Middle panel: evolution of UV (rest-frame 1500 \AA) luminosity, 
$L_\lambda ({\rm 1500 \AA})$. 
Bottom panel: evolution of $Q_{\rm H}/L_\lambda ({\rm 1500 \AA})$ normalized 
by that for $Z=Z_\odot$. 
The meaning of each curve is the same as that in Figure \ref{fig:sfrd}. 
\label{fig:QLratio}
}
\end{figure}

The upper panel of Figure \ref{fig:QLratio} shows the evolution of ionizing photon production rate 
($Q_{\rm H}$) for a constant star formation rate (${\rm SFR}=1M_\odot{\rm yr^{-1}}$) 
calculated by Schaerer (2003). 
It is shown that the ionizing photon production rate becomes larger with 
decreasing metallicity and with increasing upper mass limit. 
The middle panel of Figure \ref{fig:QLratio} shows the evolution of UV (rest-frame 1500 \AA) luminosity for the same models 
as the upper panel of Figure \ref{fig:QLratio}. For $\log t < 6.5$ the UV luminosity becomes fainter with decreasing 
metallicity. This occurs because the effective temperature 
becomes hotter with decreasing metallicity. 
Since the ionizing photon production rate density is converted from the UV luminosity 
density, the dependence of $Q_{\rm H}/L_\lambda({\rm 1500 \AA})$ on $Z$ is 
important to discuss the ionizing photon production rate 
density in high-$z$ universe. 
Figure \ref{fig:QLratio} shows the evolution of $Q_{\rm H}/L_\lambda({\rm 1500 \AA})$ normalized 
by that for $Z=Z_\odot$ as a function of metallicity. 
For galaxies with $Z \ge 0.02Z_\odot$, the ratio is smaller than 1.82. 
On the other hand, for the case of Population III, the ratio is larger than 2.35 
and can reach 4.58 (5.19) for Salpeter IMF with 
$M_{\rm up}=100M_\odot$($500M_\odot$).

\clearpage

{\footnotesize
\begin{deluxetable}{cccccc}
\tablecaption{\label{tab:candidats}Candidate Galaxies at $z \sim 8$ in the HUDF}
\tablewidth{0pt}
\tablehead{
\colhead{No.} &
\colhead{Bouwens ID} & 
\colhead{Bunker ID} & 
\colhead{McLure ID} & 
\colhead{Yan ID} & 
\colhead{Finkelstein ID} 
}
\startdata
  1& y-37636015 & YD7   & 2079  & z8-B114 &  200  \\
  2& y-37796000 & YD2   & 1939  & z8-B117 &  213  \\
  3& y-38135539 & YD3   & 1721  & z8-B115 &  125  \\
  4& y-42886345 & YD1   & 1765  & z8-B092 &  819  \\
  5& y-43086276 &\nodata& 2841  & z8-B088 &  653  \\
  6& \nodata    & YD4   & 2487  & \nodata &\nodata\\
  7& \nodata    & YD5   &\nodata& z8-SD24 &\nodata\\
  8& \nodata    &\nodata& 1422  & \nodata & 2055  \\
\hline
  9& \nodata    & YD6   &\nodata& \nodata &\nodata\\
 10& \nodata    &\nodata&\nodata& z8-B041 &\nodata\\
 11& \nodata    &\nodata&\nodata& z8-B094 &\nodata\\
 12& \nodata    &\nodata&\nodata& z8-B087 &\nodata\\
 13& \nodata    &\nodata&\nodata& z8-SB27 &\nodata\\
 14& \nodata    &\nodata&\nodata& z8-SB30 &\nodata\\
 15& \nodata    &\nodata&\nodata& z8-SD05 &\nodata\\
 16& \nodata    &\nodata&\nodata& z8-SD02 &\nodata\\
 17& \nodata    &\nodata&\nodata& z8-SD52 &\nodata\\
 18& \nodata    &\nodata&\nodata& z8-SD15 &\nodata\\
 19& \nodata    &\nodata&\nodata& \nodata &  800  \\
 20& \nodata    &\nodata&\nodata& \nodata &  640  \\
\hline
 21\tablenotemark{*}& z-44716442\tablenotemark{**}& zD7 & 1107 & z7-A044 & 3022 \\
\enddata

\tablenotetext{*}{
Although the galaxy No.21 is selected as $z \sim 8$ 
galaxy from  the photometric redshift (McLure et al. 2010; Finkelstein et al. 2010), 
it is selected as $z \sim 7$ galaxy from the dropout method (Oesch et al. 2010; Bunker et al 2010; 
Yan et al. 2010). In this paper, we categorize this galaxy as $z \sim 7$ galaxy.}
\tablenotetext{**}{Oesch et al. (2010).}

\end{deluxetable}
}

{\footnotesize
\begin{deluxetable}{lccc}
\tablecaption{\label{tab:selection}Summary of Selection Criteria of $z \sim 8$ Galaxies.}
\tablewidth{0pt}
\tablehead{
\colhead{Reference} & 
\colhead{$Y_{105}-J_{125}$ vs. $J_{125}-H_{160}$} &
\colhead{ACS Bands} &
\colhead{$J_{125}$ (mag)} 
}
\startdata
Bouwens et al. (2010b) & $Y_{105}-J_{125}>0.8$, $J_{125}-H_{160}<0.5$, & No detection ($<2 \sigma$) in all bands & $\le 28.6$ \\
               & and $(J_{125}-H_{160}) < 0.2 + 0.12(Y_{105}-J_{125})$ & & \\
Bunker et al. (2010) & $Y_{105}-J_{125}>1$ & No detection ($< 2\sigma$ ) in all bands & $<28.5$\\
McLure et al. (2010) & \multicolumn{2}{c}{photometric redshift} & \\
Yan et al.    (2010) & $Y_{105}-J_{125} > 0.8$ & No detection ($< 2\sigma$) in all bands &  $\le 29.0$ \\
Finkelstein et al. (2010) & \multicolumn{2}{c}{photometric redshift} & \\
\enddata                                                                                  

\end{deluxetable}
}

\begin{deluxetable}{llllllll}
\scriptsize
\tablecaption{\label{tab:readz8} Summary of Photometric Properties of Our Robust Sample of Galaxies at $z \sim 8$ }
\tablewidth{0pt}
\tablehead{
\colhead{{\#}} &
\colhead{ref.} & 
\colhead{Bouwens} & 
\colhead{Bunker} & 
\colhead{McLure} & 
\colhead{Yan} & 
\colhead{Finkelstein} & 
\colhead{ave.} 
}
\startdata
1 & $J$   & $28.6 \pm 0.3$ & $28.44 \pm 0.16$ & $29.01 \pm 0.38$ & $28.94 \pm 0.23$ & $28.02 \pm 0.12$ & $28.60 \pm 0.42$ \\
    & $Y-J$ & $>1.8$         & $1.13 \pm 0.14$  & $0.61 \pm 0.79$  & $>0.56$          & $>1.03$          & $0.87 \pm 0.54$  \\
    & $J-H$ & $0.0 \pm 0.3$  & $-0.17 \pm 0.24$ & $-0.32 \pm 0.63$ & $-0.55 \pm 0.43$ & $-0.27 \pm 0.18$ & $-0.26 \pm 0.27$ \\
\hline
2 & $J$   & $27.9 \pm 0.1$ & $27.88 \pm 0.10$ & $28.33 \pm 0.23$ & $28.49 \pm 0.19$ & $28.70 \pm 0.17$ & $28.26 \pm 0.37$ \\
    & $Y-J$ & $1.6 \pm 0.9$  & $2.24 \pm 0.73$  & $>1.67$          & $>1.02$          & $>0.62$          & $1.92 \pm 0.74$  \\
    & $J-H$ & $-0.3 \pm 0.2$ & $-0.19 \pm 0.14$ & $-0.16 \pm 0.35$ & $0.09 \pm 0.25$  & $-0.07 \pm 0.25$ & $-0.13 \pm 0.18$ \\
\hline
3 & $J$   & $28.4 \pm 0.2$ & $28.07 \pm 0.11$ & $28.41 \pm 0.24$ & $28.39 \pm 0.17$ & $28.61 \pm 0.14$ & $28.38 \pm 0.21$ \\
    & $Y-J$ & $>2.1$         & $1.70 \pm 0.54$  & $>1.59$          & $>1.13$          & $>0.86$          & $1.70 \pm 0.54$  \\
    & $J-H$ & $0.1 \pm 0.2$  & $0.00 \pm 0.15$  & $0.25 \pm 0.31$  & $0.03 \pm 0.22$  & $0.30 \pm 0.18$  & $0.14 \pm 0.17$  \\
\hline
4 & $J$   & $28.1 \pm 0.2$ & $27.70 \pm 0.08$ & $28.23 \pm 0.21$ & $28.45 \pm 0.18$ & $28.09 \pm 0.12$ & $28.11 \pm 0.28$ \\
    & $Y-J$ & $1.1 \pm 0.4$  & $1.19 \pm 0.25$  & $1.21 \pm 0.64$  & $0.85 \pm 0.55$  & $0.73 \pm 0.30$  & $1.02 \pm 0.30$  \\
    & $J-H$ & $-0.1 \pm 0.2$ & $-0.28 \pm 0.14$ & $0.12 \pm 0.29$  & $-0.08 \pm 0.25$ & $-0.02 \pm 0.17$ & $-0.07 \pm 0.17$ \\
\hline
5 & $J$   & $28.6 \pm 0.2$ & \nodata          & $28.98 \pm 0.37$ & $28.86 \pm 0.22$ & $28.31 \pm 0.15$ & $28.69 \pm 0.32$ \\
    & $Y-J$ & $1.1 \pm 0.5$  & \nodata          & $>1.02$          & $>0.67$          & $>0.73$          & $1.10 \pm 0.50$    \\
    & $J-H$ & $-0.6 \pm 0.3$ & \nodata          & $-0.43 \pm 0.65$ & $-0.53 \pm 0.40$ & $-0.61 \pm 0.30$ & $-0.54 \pm 0.23$ \\
\hline
6 & $J$   & \nodata        & $28.11 \pm 0.11$ & $28.59 \pm 0.28$ & \nodata          & \nodata          & $28.35 \pm 0.37$ \\
    & $Y-J$ & \nodata        & $1.63 \pm 0.67$  & $1.14 \pm 0.81$  & \nodata          & \nodata          & $1.38 \pm 0.63$  \\
    & $J-H$ & \nodata        & $-0.79 \pm 0.26$ & $-0.59 \pm 0.52$ & \nodata          & \nodata          & $-0.69 \pm 0.32$ \\
\hline
7 & $J$   & \nodata        & $28.38 \pm 0.14$ & \nodata          & $29.04 \pm 0.30$ & \nodata          & $28.71 \pm 0.50$ \\
    & $Y-J$ & \nodata        & $1.06 \pm 0.42$  & \nodata          & $>0.50$          & \nodata          & $1.06 \pm 0.42$  \\
    & $J-H$ & \nodata        & $0.41 \pm 0.17$  & \nodata          & $0.04 \pm 0.40$  & \nodata          & $0.22 \pm 0.34$  \\
\hline
8 & $J$   & \nodata        & \nodata          & $28.07 \pm 0.19$ & \nodata          & $27.58 \pm 0.11$ & $27.82 \pm 0.36$ \\
    & $Y-J$ & \nodata        & \nodata          & $0.83 \pm 0.42$  & \nodata          & $0.83 \pm 0.32$  & $0.83 \pm 0.26$  \\
    & $J-H$ & \nodata        & \nodata          & $0.24 \pm 0.25$  & \nodata          & $0.37 \pm 0.14$  & $0.30 \pm 0.17$  \\
\enddata
\end{deluxetable}

{\footnotesize
\begin{deluxetable}{cccccc}
\tablecaption{\label{tab:fesc}Minimum Escape Fraction to Ionize the Universe}
\tablewidth{0pt}
\tablehead{
    &                 &                 & \multicolumn{3}{c}{$f_{\rm esc}$} \\
$Z$ & $M_{\rm upper}$ & $M_{\rm lower}$ & $L \ge L^*$ & $L \ge 0.1 L^*$ & $L \ge 0.01 L^*$ \\
\colhead{$(Z_\odot)$} &
\colhead{$(M_\odot)$} & 
\colhead{$(M_\odot)$} & 
\colhead{} & 
\colhead{} & 
\colhead{} 
}
\startdata
\multicolumn{6}{c}{Bouwens et al. (2010b)}\\
                   &     &   & $M_{\rm UV} \le -19.5$ & $M_{\rm UV} \le -17.0$ & $M_{\rm UV} \le -14.5$ \\
\hline
$1               $ & 100 & 1 &\nodata & \nodata  & 0.85 \\
$0.02            $ & 100 & 1 &\nodata & 0.83 & 0.50 \\
$5 \times 10^{-4}$ & 100 & 1 &\nodata & 0.54 & 0.33 \\
$0               $ & 100 & 1 &\nodata & 0.31 & 0.19 \\
\hline
$5 \times 10^{-4}$ & 500 & 1 &\nodata & 0.46 & 0.28 \\
$0               $ & 500 & 1 &\nodata & 0.28 & 0.17 \\
\hline
\multicolumn{6}{c}{Bouwens et al. (2010c)}\\
                   &     &   & $M_{\rm UV} \le -20.28$ & $M_{\rm UV} \le -17.78$ & $M_{\rm UV} \le -15.28$ \\
\hline
$1               $ & 100 & 1 &\nodata & \nodata  & 0.69 \\
$0.02            $ & 100 & 1 &\nodata & 0.90 & 0.41 \\
$5 \times 10^{-4}$ & 100 & 1 &\nodata & 0.59 & 0.26 \\
$0               $ & 100 & 1 &\nodata & 0.34 & 0.15 \\
\hline
$5 \times 10^{-4}$ & 500 & 1 &\nodata & 0.50 & 0.23 \\
$0               $ & 500 & 1 &\nodata & 0.31 & 0.14 \\
\enddata
\end{deluxetable}
}

\clearpage

\clearpage

{
\epsscale{1.0}
\plotone{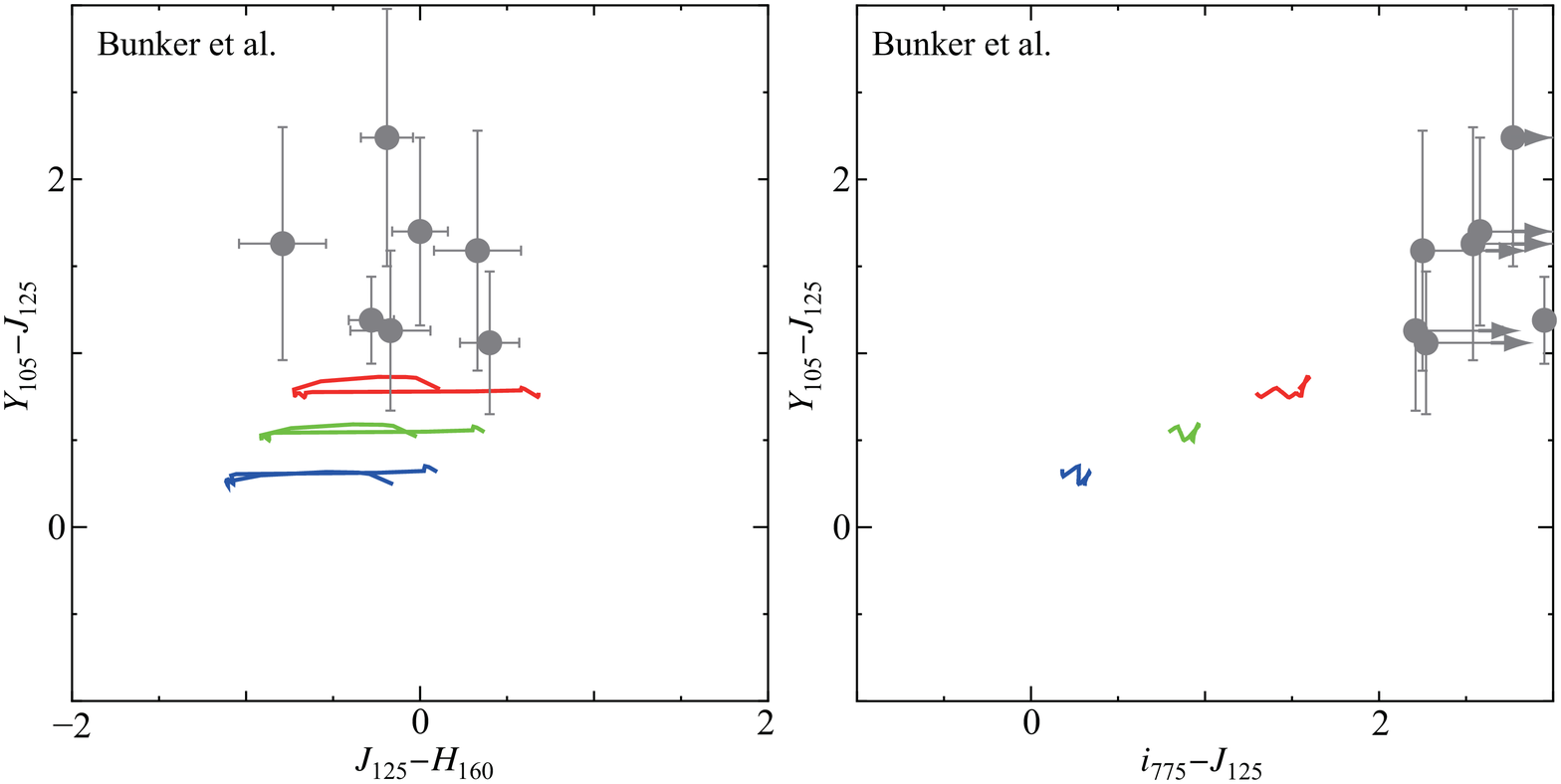}

\centerline{
Fig. 4b. --- continued.
}
}

{
\epsscale{1.0}
\plotone{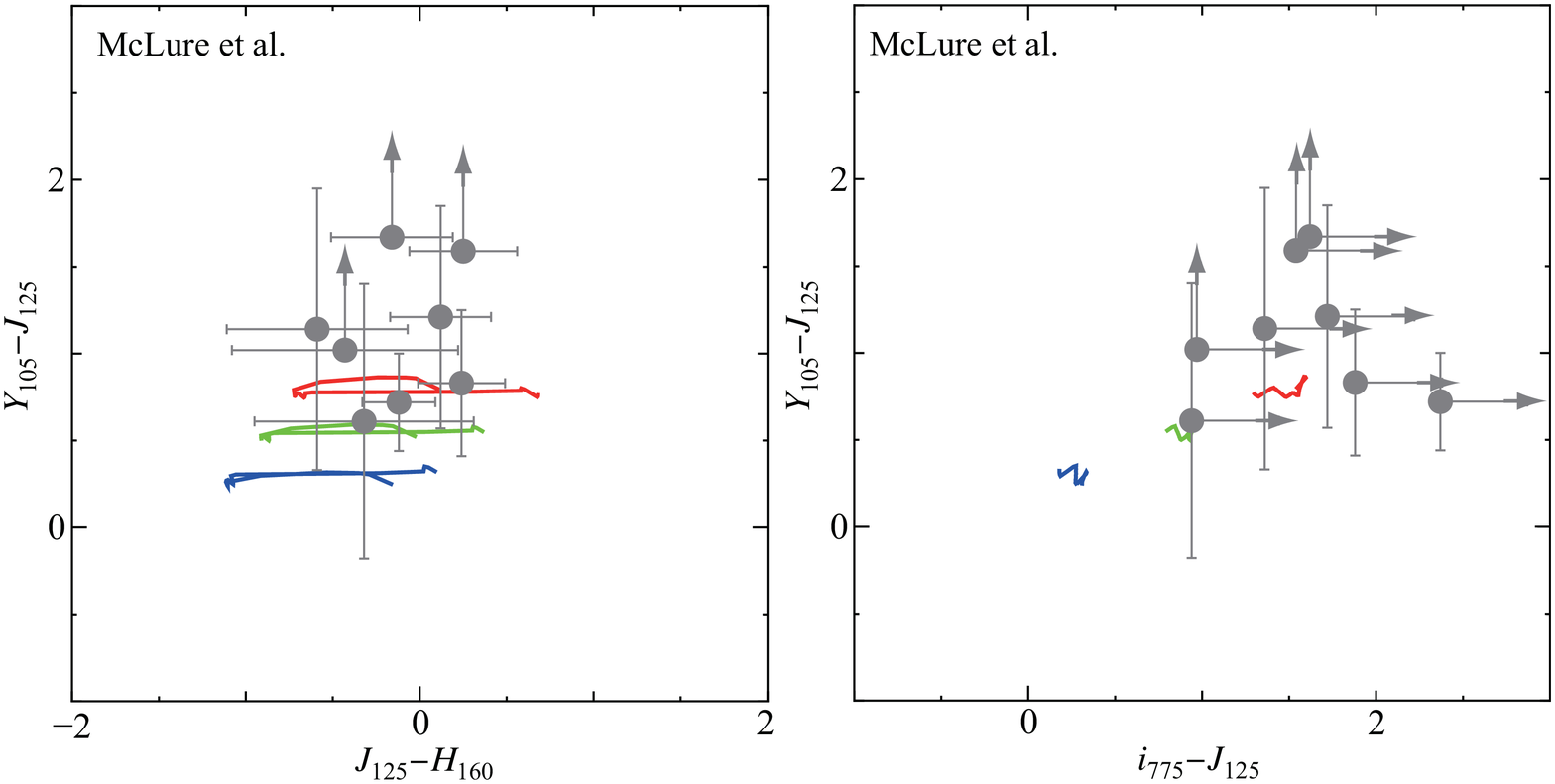}

\centerline{
Fig. 4c. --- continued.
}
}

{
\epsscale{1.0}
\plotone{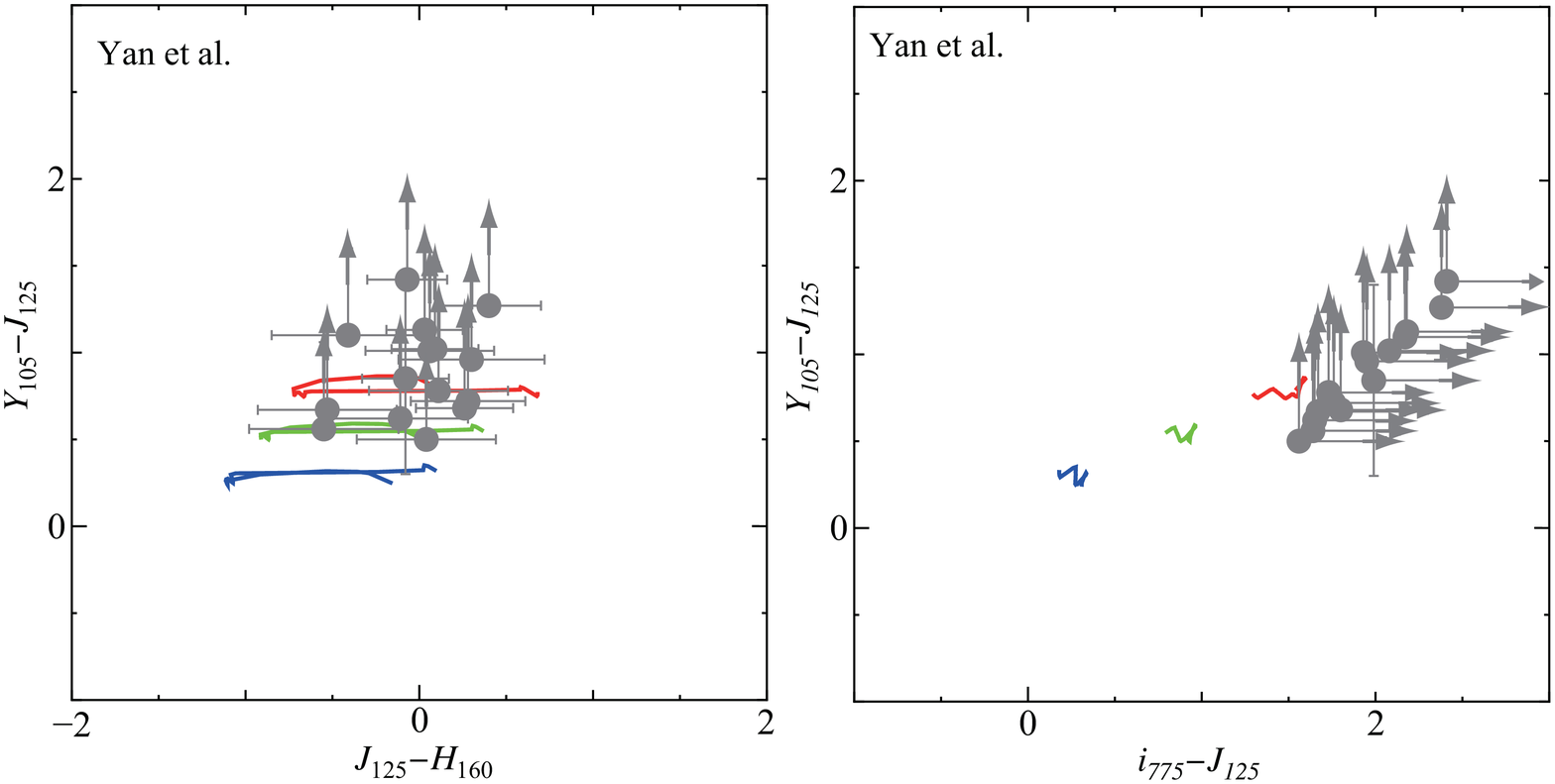}

\centerline{
Fig. 4d. --- continued.
}
}

{
\epsscale{1.0}
\plotone{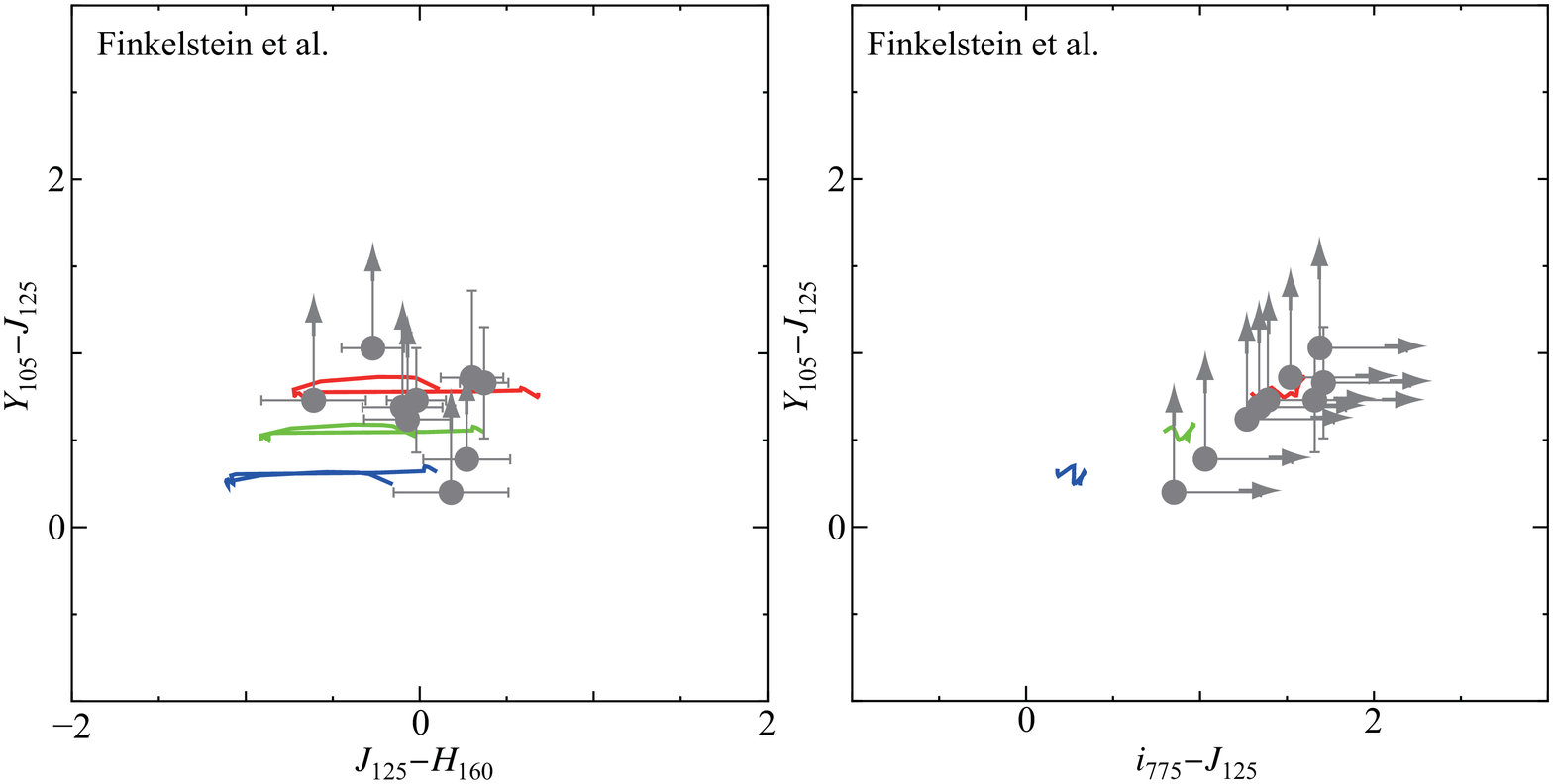}

\centerline{
Fig. 4e. --- continued.
}
}

\clearpage

\clearpage

\end{document}